\documentclass[a4paper,12pt]{article}  

\usepackage{graphicx}
\usepackage{amssymb}
\usepackage{multirow}
\usepackage{bm}
\usepackage{xcolor}
\usepackage{hyperref}

\title{Numerov and phase-integral methods for charmonium}

\date{\today}

\author{
Giampiero Esposito ORCID: 0000-0001-5930-8366 \\
Dipartimento di Fisica ``Ettore Pancini''
and INFN Sezione di Napoli, \\
Complesso Universitario di Monte S. Angelo, \\
Via Cintia Edificio 6, 80126 Napoli, Italy \and
Pietro Santorelli ORCID: 0000-0002-5882-9212 \\
Dipartimento di Fisica ``Ettore Pancini''
and INFN Sezione di Napoli, \\
Complesso Universitario di Monte S. Angelo, \\
Via Cintia Edificio 6, 80126 Napoli, Italy}

\begin{document}

\maketitle

\begin{abstract}
This paper applies the Numerov and phase-integral methods to the
stationary Schr\"{o}dinger equation that studies bound states of charm 
anti-charm quarks. The former is a numerical method
well suited for a matrix form of second-order ordinary differential 
equations, and can be applied whenever the stationary states admit
a Taylor-series expansion. The latter is an analytic method that
provides, in principle, even exact solutions of the stationary
Schr\"{o}dinger equation, and well suited for applying matched
asymptotic expansions and higher order quantization conditions.
The Numerov method is found to be always in agreement with the
early results of Eichten et al., whereas an original evaluation
of the phase-integral quantization condition clarifies under
which conditions the previous results in the literature 
on higher-order terms can be obtained. 
\end{abstract}

\section{Introduction}
\setcounter{equation}{0}
\label{sect:1}

As Dirac pointed out in his book of quantum mechanics 
\cite{Dirac}, if the exact equations
of a physical theory are too difficult, they are of no use, and one has to
resort to approximations. In particular, it is extremely difficult 
to evaluate the effective form of interquark forces from quantum chromodynamics,
and hence since the late seventies it became of interest to consider non-relativistic
models of charmed quarks. For this purpose, the work in Ref. \cite{Eichten}
assumed an instantaneous potential which is a superposition of a linear and a
Coulomb term. Since then, many original investigations of bound states for 
charmonium appeared in the literature, for which we do not even attempt to
write a comprehensive list here.

We instead remark that the resulting stationary Schr\"{o}dinger equation can
be studied with the help of advanced numerical and analytic tools, and it
has been our aim to test the efficiency of such tools for the non-relativistic
Coulomb plus linear potential model. For this purpose, Sec. \ref{sect:2} outlines
the Numerov method \cite{Numerov}, which has found so far a wide range of
applications: maximal adaptation to the Schr\"{o}dinger equation \cite{AP1},
reduction of the number of stages in the algorithm \cite{AP2},
second-order differential equations with oscillating solutions \cite{AP3},
two-point boundary-value problems \cite{AP4}, $7$ stages eighth-order methods
\cite{AP5}, vibrational eigenstates of linear triatomic molecules \cite{AP6}.
Section \ref{sect:3} studies instead the phase-integral
method \cite{FR1,FR2}, which is an improvement of the JWKB method, better 
suited for studying the one-directional nature of connection formulas and
higher-order quantization conditions. The integrals occurring in the
phase-integral quantization condition are studied in Sec. \ref{sect:4},
while numerical results are displayed in
Sec. \ref{sect:5}, concluding remarks are presented in Sec. \ref{sect:6}
and relevant details are provided in the Appendices.

\section{Numerov's method}
\setcounter{equation}{0}
\label{sect:2}

A very simple and powerful numerical approach to solve the one-dimensional (or the 
radial part of the 3-dimensional) stationary Schr\"odinger equation
is the {\it Numerov method} \cite{Blatt,Chow,RHLandau}. The method is useful to 
integrate second-order differential equations of the general form
\begin{equation}
\frac{d^2 y(x)}{dx^2} = - g(x) y(x) + s(x), 
\label{e:schr}
\end{equation}
with Cauchy data at some point $x_{0}$
\begin{equation}
y(x_0)=y_0, \hspace{1cm} y^\prime(x_0) = y_0^\prime .
\end{equation}
In the case of the stationary Schr\"odinger equation as we will discuss in the 
next section, one has
\begin{equation}
y(x) = \psi(x) \hspace{1cm} s(x) = 0, \hspace{1cm} g(x) = \frac{2m}{\hbar^{2} }(E-V(x)),
\end{equation}
where $V(x)$ is either the potential in one spatial dimension, 
or an effective potential that includes also the effects 
of angular momentum in the three-dimensional spatial case. 
Let us briefly outline the method. If the stationary state $\psi(x)$ admits
a Taylor series expansion, one can write that 
\begin{eqnarray}
\psi(x \pm \delta) & = & \psi(x) \pm \frac{d\psi(x)}{dx}\delta 
+ \frac{1}{2}  \frac{d^2\psi(x)}{dx^2}\delta^2  
\pm \frac{1}{3!}  \frac{d^3\psi(x)}{dx^3}\delta^3 
\nonumber \\
& + & \frac{1}{4!}  \frac{d^4\psi(x)}{dx^4}\delta^4 + 
\mathcal{O}(\delta^{5}),  
\end{eqnarray} 
which displays the fourth-order nature of the algorithm;
hence we can obtain the second derivative of $\psi(x)$ by considering
\begin{equation}
\frac{\psi(x+\delta) + \psi(x - \delta) - 2 \psi(x)}{\delta^2}=
\frac{d^2\psi(x)}{dx^2} +
\frac{1}{12}\frac{d^4\psi(x)}{dx^4} \delta^2 +\mathcal{O}(\delta^4).
\label{e:d2psi}
\end{equation}
Upon using Eq. (\ref{e:schr}) i.e. $\frac{d^2\psi(x)}{dx^2} =-g(x)\psi(x)$, Eq. (\ref{e:d2psi}) 
can be regarded as an equation for the fourth 
derivative of  $\psi$, giving at order $\mathcal{O}(\delta^4)$
\begin{eqnarray}
-g(x) \psi(x) & = &  \frac{\psi(x+\delta) + \psi(x - \delta) - 2 \psi(x)}{\delta^2}  
\nonumber\\
& - & 
\frac{1}{12}\bigg [ g(x+\delta)\psi(x+\delta) + g(x-\delta)\psi(x-\delta)
\nonumber \\
& - & 2 g(x)\psi(x)\bigg].
\label{e:num1}
\end{eqnarray}
If we consider a finite box (i.e. a closed interval) for the variable $x$, $[x_0,x_N]$, 
and a lattice of N points $x_i$ evenly spaced by $\delta$, Eq. (\ref{e:num1})
for $i\in\{1,...N-1\}$ can be written as
\begin{eqnarray}
\; & \; &
-\frac{\hbar^2}{2m}\frac{\psi_{i-1}-2\psi_i+\psi_{i+1}}{\delta^2}+
\frac{V_{i-1}\psi_{i-1}+ 10 V_{i}\psi_{i}+V_{i+1}\psi_{i+1}}{12} 
\nonumber \\
& = & E\frac{\psi_{i-1}+10\psi_i+\psi_{i+1}}{12},
\label{e:num2}
\end{eqnarray}
where $\psi_i \equiv \psi(x_i)$ and $V_i \equiv V(x_i)$. 
As was pointed out in Ref. \cite{Pillaietal}, 
by defining the $N$-dimensional column vector 
$\bm\psi= (...\psi_{i-1},\psi_{i},\psi_{i+1},...)$, Eq. (\ref{e:num2}) reads as
\begin{equation}
-\frac{\hbar^2}{2m} {\hat A} {\bm\psi} + {\hat B}{\hat V}{\bm \psi} = E {\hat B} {\bm \psi}
\end{equation}
where we have defined the matrices $\hat A = (\mathbb{I}_{-1} - 2\, 
\mathbb{I}_{0} +\mathbb{I}_{+1})/\delta^2$,
$\hat B = (\mathbb{I}_{-1} + 10\, \mathbb{I}_{0} +\mathbb{I}_{+1})/12$ and 
$\hat V = {\rm diag}(..., V_{i-1},V_i,V_{i+1}, ... )$ while $\mathbb{I}_{p}$ 
is the matrix of unit entries along the p$th$ diagonal  
($\mathbb{I}_{0}$ is the Identity $N\times N$ matrix). 
The matrix $\hat B$ is invertible and therefore 
\begin{equation}
-\frac{\hbar^2}{2m} {\hat B}^{-1} {\hat A} {\bm\psi} + {\hat V}{\bm \psi} = E\,  {\bm \psi}
\end{equation}
which can be viewed as an {\it eigenvalue} problem and easily solved by a 
simple numerical code, for example, with the help of Mathematica.

\section{The phase-integral method}
\setcounter{equation}{0}
\label{sect:3}

Both in one-dimensional problems and in the case of central potentials
in three-dimensional Euclidean space, the Schr\"{o}dinger equation
for stationary states leads eventually to a second-order ordinary
differential equation having the form 
\begin{equation}
\left[{d^{2}\over dz^{2}}+R(z)\right]\psi(z)=0, \; \; \;
R(z) \equiv {2m \over \hbar^{2}}(E-V(z)),
\label{(3.1)}
\end{equation}
where $V(z)$ is either the potential in one spatial dimension, or
an effective potential that includes also the effects of angular
momentum. The notation $z$ for the independent variable means that
one can study Eq. (3.1) in the complex-$z$ plane, restricting attention
to real values of $z$ only at a later stage. 
In the phase-integral method, one looks for two linearly independent,
exact solutions of Eq. (3.1) in the form
\begin{equation}
\psi(z)=D(z)e^{\pm i w(z)}.
\label{(3.2)}
\end{equation}
Since the Wronskian of two linearly independent solutions of Eq. (3.1)
is a non-vanishing constant, while the Wronskian of the functions (3.2)
is $-2iD^{2}{dw \over dz}$, for consistency one finds
\begin{equation}
D(z)={{\rm constant}\over \sqrt{dw \over dz}}.
\label{(3.3)}
\end{equation}
One can therefore write (up to a multiplicative constant)
\begin{equation}
\psi(z)={1 \over \sqrt{dw \over dz}}e^{\pm i w(z)}
={1 \over \sqrt{q(z)}} e^{\pm i \int^{z} q(\tau){\rm d}\tau},
\label{(3.4)}
\end{equation}
where $w(z) \equiv \int^{z}q(\tau)d\tau$ is said to be the
{\it phase-integral}, while $q(z)$ is the {\it phase integrand}
\cite{FR1,FR2}. By virtue of Eqs. (3.1)-(3.4), the exact phase integrand
$q(z)$ solves the differential equation
\begin{equation}
\chi(q(z)) \equiv q^{-{3 \over 2}}{d^{2}\over dz^{2}}q^{-{1 \over 2}}
+{R(z) \over q^{2}}-1=0,
\label{(3.5)}
\end{equation}
which is called the $q$-equation. Suppose now
that it is possible to determine a function 
$Q: z \rightarrow Q(z)$ that is an approximate solution of the
$q$-equation (3.5). This means that $\chi_{0}$, defined by
\begin{equation}
\chi_{0} \equiv \chi(Q(z))=Q^{-{3 \over 2}}
{d^{2}\over dz^{2}}Q^{-{1 \over 2}}
+{R(z)\over Q^{2}}-1,
\label{(3.6)}
\end{equation}
and re-expressible in the form
\begin{equation}
\chi_{0}={1 \over 16Q^{6}}\left[5 
\left({dQ^{2}\over dz}\right)^{2}-4Q^{2}
{d^{2}\over dz^{2}}Q^{2}\right]
+{R(z) \over Q^{2}}-1,
\label{(3.7)}
\end{equation}
must be much smaller than $1$. The work in Ref. \cite{FR2}
proves that the phase integrand $q(z)$ is related to the
freely specifiable base function $Q(z)$ by the asymptotic expansion
\begin{eqnarray}
q(z) & \sim & Q(z) \sum_{j=0}^{N}Y_{2j}(z) 
\nonumber \\
& \sim & Q(z) \left[1+{1 \over 2}\chi_{0}
-{1 \over 8}\chi_{0}^{2}-{1 \over 8} \left(
-{1 \over 2Q^{4}}{dQ^{2}\over dz}
{d \chi_{0}\over dz}+{1 \over Q^{2}}
{d^{2}\chi_{0}\over dz^{2}}\right)\right].
\label{(3.8)}
\end{eqnarray}
In order to obtain a stationary state that is regular at the origin, 
all choices of $Q$ fulfilling the condition \cite{FR1,FR2}
\begin{equation}
\lim_{z \to 0} z^{2} \Bigr[Q^{2}(z)-R(z) \Bigr]
=-{1 \over 4}
\label{(3.9)}
\end{equation}
are admissible. 

In our work, we have exploited precisely this freedom,
by writing $R(z)$ and $Q^{2}(z)$ in a form compatible with (3.9), 
i.e. \cite{FR1,FR2,TL1}
\begin{equation}
R(z)=A-z+{B \over z}-{l(l+1)\over z^{2}},
\label{(3.10)}
\end{equation}
\begin{equation}
Q^{2}(z)=A-z+{B \over z}-{\left(l+{1 \over 2}\right)^{2}\over z^{2}},
\label{(3.11)}
\end{equation}
where, on denoting by $a$ and $b$ the dimensionful parameters in the 
linear plus Coulomb potential
\begin{equation}
V(r)=ar-{b \over r},
\label{(3.12)}
\end{equation}
the dimensionless variables $z$, $A$ and $B$ are \cite{TL1}
\begin{equation}
z=\left({2ma \over \hbar^{2}}\right)^{1 \over 3}r, \;\;\;\;\;
A=\left({2m \over \hbar^{2}a^{2}}\right)^{1 \over 3}E, \;\;\;\;\;
B=\left({4m^{2}\over \hbar^{4}a}\right)^{1 \over 3}E,
\label{(3.13)}
\end{equation}
$E$ being the energy levels for bound states. In other words,
following Refs. \cite{TL1,L1996}, we assume that bound states with 
energy $E$ exist, and that $-Q^{2}$ has distinct zeros $x_{1}$
and $x_{2}$ on the positive half-line, and a negative zero $x_{0}$
on the negative half-line. The phase-integral quantization 
condition is \cite{FR3}
\begin{equation}
\sum_{j=0}^{N}L^{(2j+1)}=\left(s+{1 \over 2}\right)\pi, 
\; \; \; s=0,1,2,...,
\label{(3.14)}
\end{equation}
up to an error term which is studied in
detail in Ref. \cite{FR1},
the $L^{(2j+1)}$ functions being defined by the integral
\begin{equation}
L^{(2j+1)}={1 \over 2}\int_{\Gamma}Q(z)Y_{2j}(z){\rm d}z,
\label{(3.15)}
\end{equation}
where the contour $\Gamma$ encloses clockwise the positive roots
$x_{1}$ and $x_{2}$ and has a branch cut from $-\infty$ to
$x_{0}<0$ and another branch cut from $x_{1}$ to $x_{2}$.

For example, if $j=0$, we deal with
\begin{equation}
L^{(1)}={1 \over 2} \int_{\Gamma}{1 \over z}P(z) 
{{\rm d}z \over \sqrt{P(z)}},
\label{(3.16)}
\end{equation}
where $P(z)$ is the polynomial of third degree such that
\begin{eqnarray}
-P(z) & = & (z-x_{0})(z-x_{1})(z-x_{2}) 
\nonumber \\
&=& \Bigr[z^{3}-(x_{0}+x_{1}+x_{2})z^{2}
+(x_{0}x_{1}+x_{1}x_{2}+x_{2}x_{0})z
\nonumber \\
& - & x_{0}x_{1}x_{2}\Bigr].
\label{(3.17)}
\end{eqnarray}
If one defines \cite{TL1}
\begin{equation}
d^{2}=x_{2}-x_{0}, \; \; \; \; \; 
k^{2}={x_{2}-x_{1} \over x_{2}-x_{0}}, \; \; \; \; \; 
\alpha^{2}={x_{2}-x_{1} \over x_{2}},
\label{(3.18)}
\end{equation}
one can solve for $x_{0},x_{1}$ and $x_{2}$ according to
\begin{equation}
x_{0}=d^{2}\left({k^{2}\over \alpha^{2}}-1 \right), \; \; \; \; \; 
x_{1}=\left({1 \over \alpha^{2}}-1 \right)k^{2}d^{2}, \; \; \; \; \; 
x_{2}={k^{2}d^{2}\over \alpha^{2}}.
\label{(3.19)}
\end{equation}
The coefficients on the second and third line of Eq. (3.17) take 
therefore the form
\begin{equation}
x_{0}+x_{1}+x_{2}={3k^{2}d^{2}\over \alpha^{2}}-(1+k^{2})d^{2},
\label{(3.20)}
\end{equation}
\begin{equation}
x_{0}x_{1}+x_{1}x_{2}+x_{2}x_{0}={k^{2}d^{4}\over \alpha^{4}}
\Bigr[(k^{2}-\alpha^{2})(2-\alpha^{2})+k^{2}(1-\alpha^{2})\Bigr],
\label{(3.21)}
\end{equation}
\begin{equation}
x_{0}x_{1}x_{2}={k^{4}d^{6}\over \alpha^{6}}(1-\alpha^{2})
(k^{2}-\alpha^{2}).
\label{(3.22)}
\end{equation}
Thus, upon changing variable (see Appendix A for the notation on
Jacobi elliptic functions)
according to \cite{TL1,L1996}
\begin{equation}
z={k^{2}d^{2}\over \alpha^{2}}(1-\alpha^{2} {\rm sn}^{2}(u)),
\label{(3.23)}
\end{equation}
the polynomial $P(z)$ can be re-expressed, after verifying 
patiently some cancellations, in the form
\begin{equation}
P(z)=k^{4}d^{6}\,{\rm sn}^{2}(u) \left ( k^{2}{\rm sn}^{4}(u)
-(1+k^{2}){\rm sn}^{2}(u)+1\right ).
\label{(3.24)}
\end{equation}
At this stage, use is made of the identities
\begin{equation}
{\rm sn}^{2}(u)=1-{\rm cn}^{2}(u), \; \; \; \; \; 
k^{2}{\rm sn}^{2}(u)=1-{\rm dn}^{2}(u),
\label{(3.25)}
\end{equation}
and one finds eventually
\begin{equation}
P(z)=k^{4}d^{6}{\rm sn}^{2}(u) \; {\rm cn}^{2}(u) \; {\rm dn}^{2}(u),
\label{(3.26)}
\end{equation}
\begin{equation}
{{\rm d}z \over \sqrt{P(z)}}=-{2 \over d}
\left[{({\rm sn}(u))' \over {\rm cn}(u) \; {\rm dn}(u)}\right]{\rm d}u
=-{2 \over d}{\rm d}u,
\label{(3.27)}
\end{equation}
because $({\rm sn}(u))'={\rm cn}(u) \; {\rm dn}(u)$. 

On passing from the $z-$ to the $u-$ integration, the integral (3.16)
becomes proportional to the integral from 
$-K(m=k^{2})$ to $K(m=k^{2})$ of an
even function of $u$. Hence one finds eventually
\begin{equation}
L^{(1)}=2k^{2}d^{3}\alpha^{2}
\int_{0}^{K(m=k^{2})}
{{\rm sn}^{2}(u) \; {\rm cn}^{2}(u) \; {\rm dn}^{2}(u) 
\over (1-\alpha^{2} {\rm sn}^{2}(u))} {\rm d}u.
\label{(3.28)}
\end{equation}
The method developed in Ref. \cite{TL1}, that we here describe in
a more detailed way, amounts to imposing the quantization conditions 
(3.14), starting of course with $L^{(1)}$:
\begin{equation}
L^{(1)}=\left(s+{1 \over 2}\right)\pi,
\label{(3.29)}
\end{equation}
and then solving numerically the resulting transcendental equation
for $x_{2}$. On the other hand, since by assumption $Q^{2}(x_{2})=0$, 
one has also from Eq. (3.11) the equation
\begin{equation}
A=x_{2}-{B \over x_{2}}+{\left(l+{1 \over 2}\right)^{2}\over
(x_{2})^{2}}.
\label{(3.30)}
\end{equation}
On multiplying both sides of Eq. (3.30) by $(x_{2})^{2}$, one finds
\begin{equation}
(x_{2})^{3}-A(x_{2})^{2}-B x_{2}
+\left(l+{1 \over 2}\right)^{2}=0.
\label{(3.31)}
\end{equation}
This tells us that the polynomial of third degree
\begin{equation}
{\widetilde P}(z)=z^{3}-Az^{2}-Bz+\left(l+{1 \over 2}\right)^{2}
\label{(3.32)}
\end{equation}
has a root at $z=x_{2}$, and we require that $x_{0}$ and $x_{1}$ should
be the remaining two roots. We can therefore exploit Eq. (3.17) and,
upon defining $x_{0}=S-T,x_{1}=S+T$, we find
\begin{equation}
A-x_{2}=x_{0}+x_{1}=2S,
\label{(3.33)}
\end{equation}
\begin{equation}
-B=x_{0}x_{1}+x_{2}(x_{0}+x_{1})=S^{2}-T^{2}+2Sx_{2},
\label{(3.34)}
\end{equation}
\begin{equation}
\left(l+{1 \over 2}\right)^{2}=-x_{0}x_{1}x_{2}
=(T^{2}-S^{2})x_{2}.
\label{(3.35)}
\end{equation}
Equation (3.33) is solved by
\begin{equation}
S={1 \over 2}(A-x_{2})
={\left(l+{1 \over 2}\right)^{2}\over 2(x_{2})^{2}}
-{B \over 2x_{2}},
\label{(3.36)}
\end{equation}
while Eq. (3.35) yields
\begin{equation}
T^{2}=S^{2}+{\left(l+{1 \over 2}\right)^{2}\over x_{2}},
\label{(3.37)}
\end{equation}
and these solutions satisfy identically Eq. (3.34). Moreover, the terms in Eq. (3.18) 
can be now expressed in the form
\begin{equation}
d^{2}=x_{2}-S+T, \;\;\;\;\;
k^{2}={x_{2}-S-T \over x_{2}-S+T}, \;\;\;\;\;
\alpha^{2}={x_{2}-S-T \over x_{2}}.
\label{(3.38)}
\end{equation}
In light of Eqs. (3.28), (3.36)-(3.38), the transcendental equation (3.29)
can be solved numerically for $s=0,1,2,...$, $l=0,1,2,...$, and various 
choices of $B$.

More precisely, the energy levels obtained from the phase-integral method 
can be characterized by means of four numbers: the order $N$ of the phase-integral
approximation, the dimensionless strength $B$ of the Coulomb potential, the
integer $s$ in the quantization condition (3.14) and the angular momentum quantum
number $l=0,1,2,...$.

\section{The $L^{(i)}$ integrals}
\setcounter{equation}{0}
\label{sect:4}

The evaluation of the integral $L^{(1)}$ in Eq. (3.28) yields (see 
Appendix A for the notation on elliptic integrals $K,E, \Pi$, with the
understanding that $m=k^{2}$ hereafter)
\begin{eqnarray}
L^{(1)}&=& d^{3}\biggr[f_{1}(m,\alpha^{2})
{(K(m)-E(m)) \over m}
+f_{1}((1-m),{\alpha'}^{2}){E(m)\over (1-m)}
\nonumber \\
&+& {2 m(1-m)(\alpha^{2}-1) \over \alpha^{2}
{\alpha '}^{2}} \Pi(\alpha^{2},m) \biggr],
\label{(4.1)}
\end{eqnarray}
where, from Ref. \cite{TL1},
\begin{equation}
f_{1}(m,\alpha^{2})={2m[(1-m)\alpha^{4}+3m^{2}(\alpha^{2}-1)]
\over 3 \alpha^{4}},
\label{(4.2)}
\end{equation}
\begin{equation}
{\alpha'}^{2}={\alpha^{2}(1-m) \over (\alpha^{2}-m)}.
\label{(4.3)}
\end{equation}

Next, the integral $L^{(3)}$ reads as
\begin{equation}
L^{(3)}={1 \over 12 d^{3}m \alpha^{2}}
\int_{u_{0}}^{u_{0}+K(m)}
{(1-\alpha^{2}{\rm sn}^{2}(u))(1+m-3m \; {\rm sn}^{2}(u)) \over
{\rm sn}^{2}(u) \; {\rm cn}^{2}(u) \; {\rm dn}^{2}(u)}{\rm d}u,
\label{(4.4)}
\end{equation}
where, as is stressed in Ref. \cite{TL1}, $u_{0}$ is 
(see our analysis after Eq. (4.23)) a
point in the complex $u$-plane such that 
${\rm sn}(u_{0}) \not = 0$, ${\rm cn}(u_{0}) \not = 0$, 
${\rm dn}(u_{0}) \not = 0$. By exploiting the identities (3.25),
here re-expressed in the form
\begin{equation}
{\rm cn}^{2}(u)=1-{\rm sn}^{2}(u), \;
{\rm dn}^{2}(u)=1-m \; {\rm sn}^{2}(u),
\label{(4.5)}
\end{equation}
the integrand $I^{(3)}$ of Eq. (4.4) can be decomposed in partial fractions with
respect to ${\rm sn}^{2}(u)$, i.e. (setting ${\rm sn}^{2}(u)=x$ for simplicity 
of notation) 
\begin{eqnarray}
\; & \; & 12d^{3}m \alpha^{2}I^{(3)}=
{(1-\alpha^{2}x)(1+m-3m x)\over x(1-x)(1-mx)}
\nonumber \\
&=& {F_{1}\over x}+{F_{2}\over (1-x)}
+{F_{3}\over (1-mx)}
\nonumber \\
&=& {F_{1}(1-x)(1-mx)+F_{2}x(1-mx)+F_{3}x(1-x) \over
x(1-x) (1-mx)},
\label{(4.6)}
\end{eqnarray}
where $F_{i}$ is a function of $m$ and $\alpha^{2}$, $\forall i=1,2,3$.
By comparison of left- and right-hand side, we find therefore the three 
equations
\begin{equation}
1+m=F_{1},
\label{(4.7)}
\end{equation}
\begin{equation}
-3m-\alpha^{2}-\alpha^{2}m=-(m+1)F_{1}+F_{2}+F_{3},
\label{(4.8)}
\end{equation}
\begin{equation}
3 \alpha^{2}m=(F_{1}-F_{2})m-F_{3}.
\label{(4.9)}
\end{equation}
By insertion of Eq. (4.7) into Eq. (4.8), and subsequent addition of Eqs. (4.8)
and (4.9), we find
\begin{equation}
F_{2}={(1-\alpha^{2})(1-2m) \over (1-m)}.
\label{(4.10)}
\end{equation}
Eventually, we obtain $F_{3}$ from Eqs. (4.7)-(4.10) in the form
\begin{equation}
F_{3}={(\alpha^{2}-m)m(m-2)\over (1-m)}.
\label{(4.11)}
\end{equation}
At this stage, we can exploit the three indefinite integrals \cite{BRY}
\begin{equation}
Z_{1}(u,m)=\int {{\rm d}u \over {\rm sn}^{2}(u)}
=-{{\rm cn}(u) \; {\rm dn}(u) \over {\rm sn}(u)}+u-{\cal E}(u,m),
\label{(4.12)}
\end{equation}
\begin{equation}
Z_{2}(u,m)=\int {{\rm d}u \over {\rm cn}^{2}(u)}
={1 \over (1-m)}{{\rm dn}(u) \; {\rm sn}(u) \over {\rm cn}(u)}
+u-{1 \over (1-m)} {\cal E}(u,m),
\label{(4.13)}
\end{equation}
\begin{equation}
Z_{3}(u,m)=\int {{\rm d}u \over {\rm dn}^{2}(u)}
=-{m\over (1-m)}
{{\rm cn}(u) \; {\rm sn}(u) \over {\rm dn}(u)}
+{1 \over (1-m)} {\cal E}(u,m),
\label{(4.14)}
\end{equation}
where the function ${\cal E}$ is defined by
\begin{equation}
{\cal E}(u,m)=\int_{0}^{u}{\rm dn}^{2}(z,m){\rm d}z.
\label{(4.15)}
\end{equation}
By virtue of Eqs. (4.6)-(4.15) we find
\begin{eqnarray}
12d^{3}m \alpha^{2}L^{(3)}&=&
(1+m)\Bigr[Z_{1}(u_{0}+K(m),m)-Z_{1}(u_{0},m)\Bigr]
\nonumber \\
&+& {(1-\alpha^{2})(1-2m)\over (1-m)}
\Bigr[Z_{2}(u_{0}+K(m),m)-Z_{2}(u_{0},m)\Bigr]
\nonumber \\
&+& {(\alpha^{2}-m)m(m-2) \over (1-m)}
\Bigr[Z_{3}(u_{0}+K(m),m)-Z_{3}(u_{0},m)\Bigr]
\nonumber \\
&=& {\cal A}(m,\alpha^{2})E(m)+{\cal B}(m,\alpha^{2})K(m)
+C(u_{0},m,\alpha^{2}),
\label{(4.16)}
\end{eqnarray}
where 
\begin{eqnarray}
{\cal A}(m,\alpha^{2})&=& -(1+m)
-{(1-\alpha^{2})(1-2m)\over (1-m)^{2}}
+{(\alpha^{2}-m)m(m-2)\over (1-m)^{2}}
\nonumber \\
&=& -{[2m^{3}-(\alpha^{2}+3)m^{2}+(4\alpha^{2}-3)m+2-\alpha^{2}]
\over (1-m)^{2}},
\label{(4.17)}
\end{eqnarray}
\begin{eqnarray}
{\cal B}(m,\alpha^{2})&=& (1+m)+{(1-\alpha^{2})(1-2m)\over (1-m)}
\nonumber \\
&=&{{-m^{2}+2(\alpha^{2}-1)m+2-\alpha^{2}}\over (1-m)},
\label{(4.18)}
\end{eqnarray}
while, upon exploiting the identity (we do not write
the second argument (i.e., $m$) of Jacobi
elliptic functions for simplicity of notation)
\begin{equation}
{\cal E}(u_{0}+K(m),m)-{\cal E}(u_{0},m)
=E(m)-m {{\rm cn}(u_{0}) \; {\rm sn}(u_{0})\over {\rm dn}(u_{0})},
\label{(4.19)}
\end{equation}
we find from Eqs. (4.12)-(4.16) and Appendix B that
\begin{eqnarray}
C(u_{0},m,\alpha^{2}) &=&
-{(m^{2}-1){\rm sn}(u_{0})\over {\rm cn}(u_{0}) \; {\rm dn}(u_{0})}
+(1+m){{\rm cn}(u_{0}) \; {\rm dn}(u_{0})\over {\rm sn}(u_{0})}
\nonumber \\
&-& {(1-\alpha^{2})(1-2m)\over (1-m)^{2}}
\left[{{\rm cn}(u_{0})\over {\rm dn}(u_{0}) \; {\rm sn}(u_{0})}
+{{\rm dn}(u_{0}) \; {\rm sn}(u_{0})\over {\rm cn}(u_{0})}\right]
\nonumber \\ 
&+&{[m^{3}-3m^{2}+2m+m(m^{2}-1)\alpha^{2}]\over (1-m)^{2}}
{{\rm cn}(u_{0}) \; {\rm sn}(u_{0}) \over {\rm dn}(u_{0})}
\nonumber \\
&=& {F(u_{0},m,\alpha^{2})\over 
{\rm cn}(u_{0}) \; {\rm dn}(u_{0}) \; {\rm sn}(u_{0})}
+G(m,\alpha^{2}){{\rm cn}(u_{0}) \; {\rm sn}(u_{0}) \over
{\rm dn}(u_{0})},
\label{(4.20)}
\end{eqnarray}
where
\begin{eqnarray}
F(u_{0},m,\alpha^{2})
&=& (m^2+m) {\rm cn}^4 (u_{0}) -m^2+1
\nonumber \\
&+& \frac{\left(\alpha ^2-1\right) \left(2 m-1\right) 
\left(m \; {\rm sn}^4(u_{0}) -1\right)}
{\left(1-m\right)^2} , 
\label{(4.21)}
\end{eqnarray}
by repeated application of Eq. (4.5), while
\begin{equation}
G(m,\alpha^{2})={\gamma(m,\alpha^{2})\over (1-m)^{2}},
\label{(4.22)}
\end{equation}
having set
\begin{eqnarray}
\gamma(m,\alpha^{2})&=&
2m^{4}-3m^{3}-3m^{2}+2m+\alpha^{2}(-m^{3}+4m^{2}-m) 
\nonumber \\
&+& (2m^{3}-4m^{2})(\alpha^{2}-m)
\nonumber \\
&=& m^{3}-3m^{2}+2m+m(m^{2}-1)\alpha^{2}.
\label{(4.23)}
\end{eqnarray}

A crucial remark is now in order. Since $u_{0}$ is in general complex, 
the quantization condition (3.14), written here in the approximate form
(cf. Ref. \cite{Froman1966})
\begin{equation}
L^{(1)}+L^{(3)}=\left(s+{1 \over 2}\right)\pi,
\label{(4.24)}
\end{equation}
would lead to complex energy eigenvalues because 
$C(u_{0},m,\alpha^{2})$ would be complex-valued, leading in turn
to complex values of $L^{(3)}$. Of course, this is inconsistent. 
As far as we can see, the only way out lies in looking for the particular
values of $u_{0}$ such that
\begin{equation}
C(u_{0},m,\alpha^{2})=0.
\label{(4.25)}
\end{equation}
By virtue of Eqs. (4.20)-(4.23), Eq. (4.25) reads as
\begin{equation}
{F(u_{0},m,\alpha^{2}) \over G(m,\alpha^{2})}
=-{\rm cn}^{2}(u_{0}) \; {\rm sn}^{2}(u_{0}).
\label{(4.26)}
\end{equation}
By virtue of Eq. (4.5), our Eq. (4.26) can be re-expressed as an
algebraic equation of second degree in the variable $x={\rm sn}^{2}(u_{0})$,
after multiplying both sides by $G(m,\alpha^{2})$ and computing patiently
all coefficients of $x^{2},x$ and $x^{0}$. Hence we find
\begin{equation}
\kappa_{2}x^{2}+\kappa_{1}x+\kappa_{0}=0,
\label{(4.27)}
\end{equation}
where
\begin{equation}
\kappa_{2}=m^{4}-2m^{3}+(2-m)m^{2} \alpha^{2},
\label{(4.28)}
\end{equation}
\begin{equation}
\kappa_{1}=-2m^{4}+3m^{3}-m^{2}+m(m^{2}-1)\alpha^{2},
\label{(4.29)}
\end{equation}
\begin{equation}
\kappa_{0}=m(m^{2}-m+1)+(1-2m)\alpha^{2}.
\label{(4.30)}
\end{equation}
We obtain therefore
\begin{equation}
x(m,\alpha^{2})={{-\kappa_{1} \pm \sqrt{(\kappa_{1})^{2}-4 \kappa_{0}\kappa_{2}}}
\over 2 \kappa_{2}},
\end{equation}
while $u_{0}$ solves the transcendental equation
\begin{equation}
{\rm sn}(u_{0},m)=\sqrt{x}.
\label{(4.32)}
\end{equation}
A particular solution can be obtained upon setting $m=1$, which
implies, from Eqs. (4.28)-(4.30),
\begin{equation}
\kappa_{2}(m=1)=\alpha^{2}-1, \;
\kappa_{1}(m=1)=0, \;
\kappa_{0}(m=1)=1-\alpha^{2},
\label{(4.33)}
\end{equation}
and hence
\begin{equation}
x(m=1)= \sqrt{-{\kappa_{0}\over \kappa_{2}}} = \pm 1,
\label{(4.34)}
\end{equation}
which leads in turn to
\begin{equation}
{\rm sn}(u_{0},m=1)= \sqrt{\pm 1}= \pm 1, \; \pm i.
\label{(4.35)}
\end{equation}
The roots $\pm 1$ in Eq. (4.35) should be ruled out, in order to
be consistent with the definition (4.4), where it is assumed that 
${\rm cn}(u_{0}) \not =0$. We can therefore consider only the equation
\begin{equation}
{\rm sn}(u_{0},m=1)=i,
\label{(4.36)}
\end{equation}
which is solved by
\begin{equation}
u_{0}=i {\pi \over 4}.
\label{(4.37)}
\end{equation}
Moreover, for generic values of $m$ and $\alpha$ we find from Eqs.
(4.31) and (4.32) the two families of solutions which formally can be written as
\begin{equation}
u_0 = {\rm sn}^{-1}\left(
\sqrt{{{-\kappa_{1} \pm \sqrt{(\kappa_{1})^{2}-4 \kappa_{0}\kappa_{2}}} \over 2 \kappa_{2}}},m
\right).
\label{(4.38)}
\end{equation}
For these values of $u_0$ our expression of $L^{(3)}$, as already discussed, is 
the same as the one in Ref. \cite{TL1}, but before performing the numerical calculations 
we should verify that these points in the complex $u$-plane are such that 
${\rm sn}(u_{0}) \not = 0$, ${\rm cn}(u_{0}) \not = 0$, 
${\rm dn}(u_{0}) \not = 0$. If one bears in mind the 
expressions in Eqs. (\ref{(3.36)})-(\ref{(3.38)}),
and Eq. (\ref{(4.38)}), it is clear that the exact location of $u_0$ in the complex plane 
can be obtained only after the evaluation of the quantization condition in Eq. (\ref{(4.24)}).
Thus, in the next section, after the calculation of the energy values (i.e. $A$ in Eq.
(\ref{(3.13)})) we will verify the consistency of the numerical analysis by verifying that the
$u_0$ values are such that  ${\rm sn}(u_{0}) \not = 0$, ${\rm cn}(u_{0}) \not = 0$ 
and ${\rm dn}(u_{0}) \not = 0$.

\section{Numerical calculations}
\label{sect:5}
\setcounter{equation}{0}

In this section we perform a numerical comparison between the  
solution of the radial part of the 3-dimensional Schr\"odinger 
equation with a central potential by using the  {\it Numerov method}  
(cf. Sec. \ref{sect:2}) and the outcomes of the phase-integral 
method discussed in the previuos sections.

In order to test our code, in Table \ref{t:Numerov} we have collected  
dimensionless energy eigenvalues 
$A$ (cf. Eq. (\ref{(3.13)})) for the parameters 
$B$ and $\ell$ occurring in the central potential discussed in Ref. \cite{TL1} 
for different mesh values. As one can see, also for a small grid
the results are in good agreement with the final results, which are obtained for $N=5000$.
In all calculations we use $[10^{-4},50]$ as the range for 
the dimensionless $z$ parameter 
(cf. (\ref{(3.13)})).  In Table $2$,  for the same set of values of 
$B$ and $\ell$ in Table \ref{t:Numerov}, we write 
the dimensionless energy eigenvalues $A$ (cf. Eq. (\ref{(3.13)})). 
The agreement is very good for low values of $B$
but decreases for large $B$, for which 
however there is an improvement 
when the value of $\ell$ increases. It seems to us that 
this last feature can be understood upon bearing in mind that, at
large $l$, the functions $R$ and $Q^{2}$ in Eqs. (3.10) and (3.11)
take almost equal values. 
More precisely, there is not a unique way of fulfilling the limiting
condition (3.9). Once $R(z)$ is given as in Eq. (3.10), our choice of
$Q^{2}(z)$, inspired by the work of Refs. \cite{FR1,FR2}, is compatible
with (3.9), but one might consider as well a choice of $Q^{2}(z)$ in
the form
\begin{equation}
Q^{2}(z)=A-z+{B \over z}-{\left(l+{1 \over 2}\right)^{2}\over z^{2}}
+f(z),
\label{(5.1)}
\end{equation}
where $f$ is any function such that 
$\lim_{z \to 0}z^{2}f(z)=0$, e.g., $f(z)=\alpha \; e^{-z}$.
This is not a rigorous argument either, but it displays clearly that
the arbitrariness in the choice of $Q^{2}(z)$ may account for different
theoretical estimates of the energy eigenvalues, since the condition of
vanishing $Q^{2}$ would no longer lead to Eq. (3.30), but rather to
the equation
\begin{equation}
A=x_{2}-{B \over x_{2}}+{\left(l+{1 \over 2}\right)^{2}\over (x_{2})^{2}}
-f(x_{2}),
\label{(5.2)}
\end{equation}
if Eq. (5.1) is taken to hold. In other words, different values of $B$
and different choices of $f(z)$ will affect the theoretical estimate
of dimensionless eigenvalues $A$.

\renewcommand{\arraystretch}{1.35}
\begin{table}
\begin{center}
\begin{tabular}{|c|c|c|c|c|c|c|c|}
\hline
\multirow{2}{*}{$(B,\ell)$}        &   \multicolumn{7}{|c|}{$A_N$}                                  \\ 
\cline{2-8}
                     &    $N=8$          &        $N=16$     &    $N=32$    &        $N=64$   &   $N=128$  &   $N=256$  &   $N=512$   \\
\hline
\hline
 (0,0)             &  2.8858    &    2.3509    
&    2.3380    &    2.3381    
&    2.3381    &    2.3381
&    2.3381     \\
 \hline
  (0,1)           &  3.2036    &    3.2472
&    3.3499   &    3.3599.  &    3.3611    &    3.3612
&    3.3613      \\
 \hline
 (0,2)            &  3.8374    &    4.2366   &    4.2471
   &    4.2481   &    4.2482 &    4.2482    &    4.2482     \\
\hline\hline
 (2,0)             &  2.0887    &    0.90968   &    0.46478    &    0.28595
&    0.22185   &    0.20221
&    0.19676       \\
 \hline
  (2,1)           &  2.4071   &    1.9879
&    2.1960   &    2.2326
&    2.2375    &    2.2381 &    2.2381       \\
 \hline
 (2,2)            &  3.0427    &    3.4096 &    3.4299    &    3.4316 &    3.4317  &    3.4317 &    3.43174     \\
\hline\hline
 (5,0)             &  0.89188   &    - 1.3705    &    1.1344. &    0.72893    &    0.53979 &    0.46719   &    0.44397     \\
 \hline
  (5,1)            & 1.2107   &    - 0.17807   &    - 0.13662   &    0.042615   &    0.075806    &    0.080476     &    0.081094   \\
 \hline
 (5,2)            & 1.8479   &    1.8685  &    2.0218    &    2.0266 &    2.0269   &    2.0269 &    2.0269    \\
\hline\hline
 (10,0)             &   -1.1047   &    0.045198 &    1.0639   &    0.30229   &    - 0.10734   &    - 0.29961   &    - 0.37306        \\
 \hline
  (10,1)           &    -0.78537    &    0.37281   &    - 0.94293    &    - 0.82315    &    - 0.63611   
  &    - 0.60284   &    - 0.59819         \\
 \hline
 (10,2)            &  -0.14698    &    1.0765   &    - 0.99667    &    - 0.94581   &    - 0.94363   &    - 0.94350   &    - 0.94349     \\
\hline\hline
\end{tabular}
\caption{\small \em In this Table we collect the results for 
$A$ (cf. Eq. (\ref{(3.13)})) 
in the Numerov ($A_{N}$) method for different mesh values. 
Note that in Table \ref{t:confronti}
the values of $A_N$ have been obtained for $N=5000$; moreover, 
our range for $z$ (cf.  Eq. (\ref{(3.13)})) is $[10^{-5},20]$.}
\label{t:Numerov}
\end{center}
\end{table}

\renewcommand{\arraystretch}{1.3}
\begin{table}

\begin{center}
\begin{tabular}{|c|c|c||c|}
\hline
$B$        &   $\ell$    & $A_{N}$        &  $A_{PhI}$       \\
              &               &                       &        $n=0 $       \\

\hline
 0           &     0        &  2.33811         & 2.34966              \\ 
 \hline
 0           &     1       &   3.36125        &  3.36536              \\
 \hline
 0           &     2       &   4.24818        & 4.25046             \\
\hline\hline
 2           &     0        &  0.194971         &  0.151574      \\
 \hline
 2           &     1       &   2.23816          &  2.23556           \\
 \hline
 2           &     2       &   3.43174        &    3.4322            \\
\hline\hline
 5           &     1       &   0.0811837       & 0.0670229           \\
 \hline
 5           &     2       &   2.02688         &  2.02359           \\
\hline\hline
 10           &     2       &  -0.943488        &  -0.952484   \\
\hline
\end{tabular}
\end{center}
\vspace{-0.5cm}
\caption{\small \em In this table we compare the results for $A$ 
(cf. Eqs. (\ref{(3.30)})  and (\ref{(3.13)})) in the Numerov ($A_{N}$) 
and in the phase-integral approach ($A_{PhI}$). We use six significant digits in 
order to make it easier to perform a comparison 
between our findings and the results in Ref. \cite{Eichten}.}
\label{t:confronti}
\end{table}

A question of crucial importance is whether the Numerov method 
displays a good convergence rate. For this purpose, following 
Refs. \cite{CONV1,CONV2,CONV3,CONV4}, one can evaluate from 
the values in Table 1 the quantity
$$
N_{k}=\log_{2}\left({|A_{k-2}-A_{k-1}|\over |A_{k-1}-A_{k}|}\right),
$$
where $A_{k}$ is, for each row, the value of $A$ displayed
in the $k$-th column.  We find, for example, for $(B=0,\ell=0)$,  
the $N_{k}$ values
$$ 
\{5.38, 6.91, 3.02, 3.87, 3.97\},
$$ 
for $(B=2,\ell=0)$ the $N_{k}$ values 
$$
\{1.41, 1.31, 1.48, 1.71, 1.85\},
$$ 
for $(B=2,\ell=2)$  the $N_{k}$ values 
$$
\{4.17, 3.61, 4.01, 4.01, 4.00\},
$$ 
and for $(B=10,\ell=2)$ the $N_{k}$ values   
$$
\{-0.761, 5.35, 4.55, 4.01, 3.99\},
$$
showing that the 
numerical convergence is quite good  for the physical cases we are considering.
Of course, as shown in Refs. \cite{CONV1,CONV2,CONV3,CONV4}, one can
also define the quantity
$$
M_{k}=\log_{2}\left({|A_{k-1}-A_{PhI}|\over |A_{k}-A_{PhI}|}\right),
$$
where $A_{PhI}$ is the value of $A$ provided by the phase-integral
method. However, in our opinion such a test merits a separate paper,
because the optimal estimate of $A$ from the phase-integral method 
is affected by different values of $B$ and of the function $f$ as we
have just discussed after Eq. (5.2).

\begin{figure}[t]
\minipage{0.5\textwidth}
\includegraphics[width=1\textwidth]{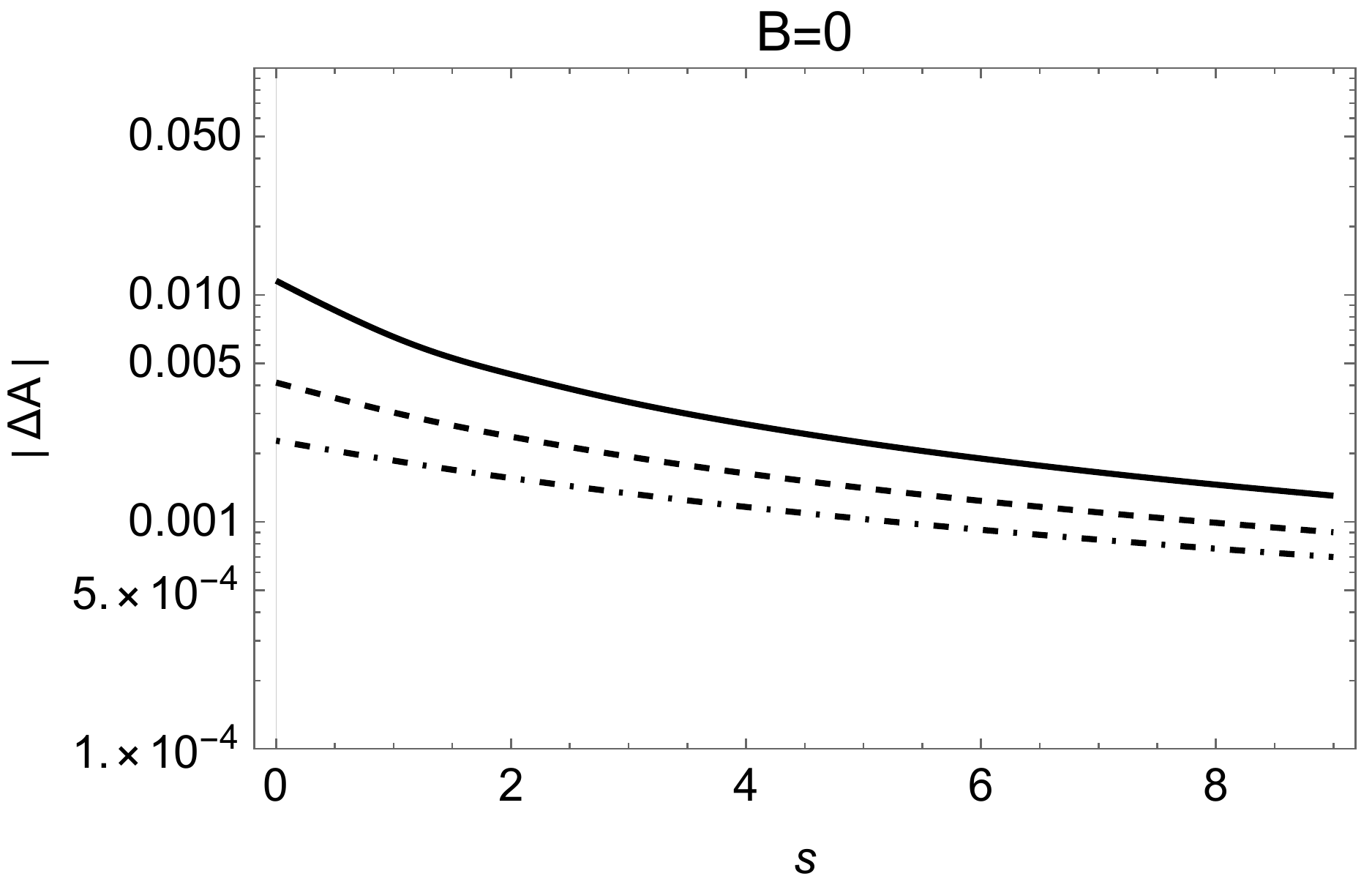}
\endminipage \hspace{0.4cm}
\minipage{0.5\textwidth}
\includegraphics[width=1\textwidth]{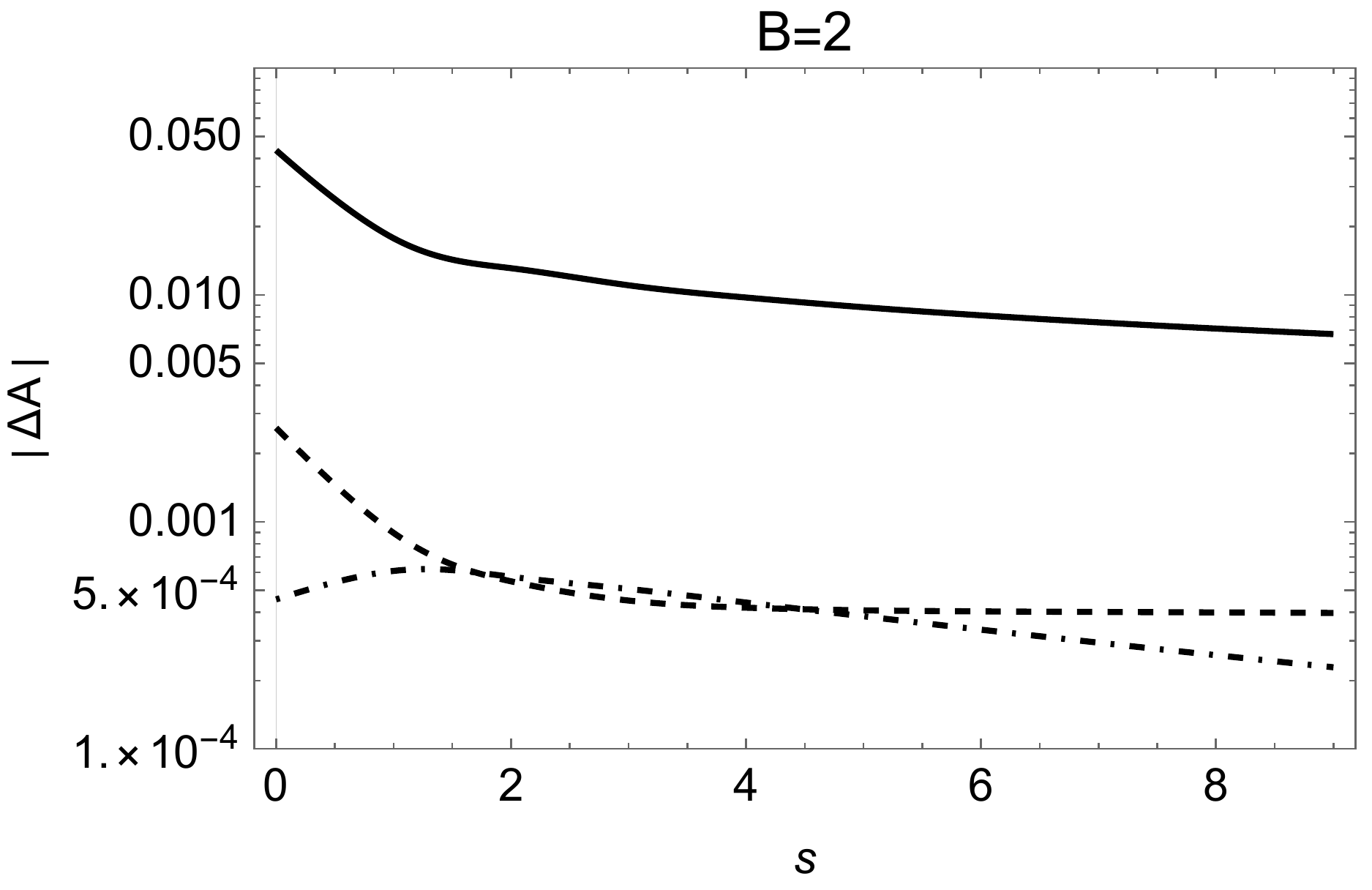}
\endminipage\\
\vspace{-0.5cm}
\caption{\small \em The $\left|\Delta A\right|$ for two different values 
of the $B$ parameter in the potential in Eq. (\ref{(3.10)}). 
The solid lines correspond to $\ell =0$, the dotted ones to $\ell =1 $ 
and the dashed-dotted lines to $\ell = 2$, respectively.  In all cases $j=0$.}
\label{fig:1}
\end{figure}

\begin{figure}[t]
\minipage{0.5\textwidth}
\includegraphics[width=1\textwidth]{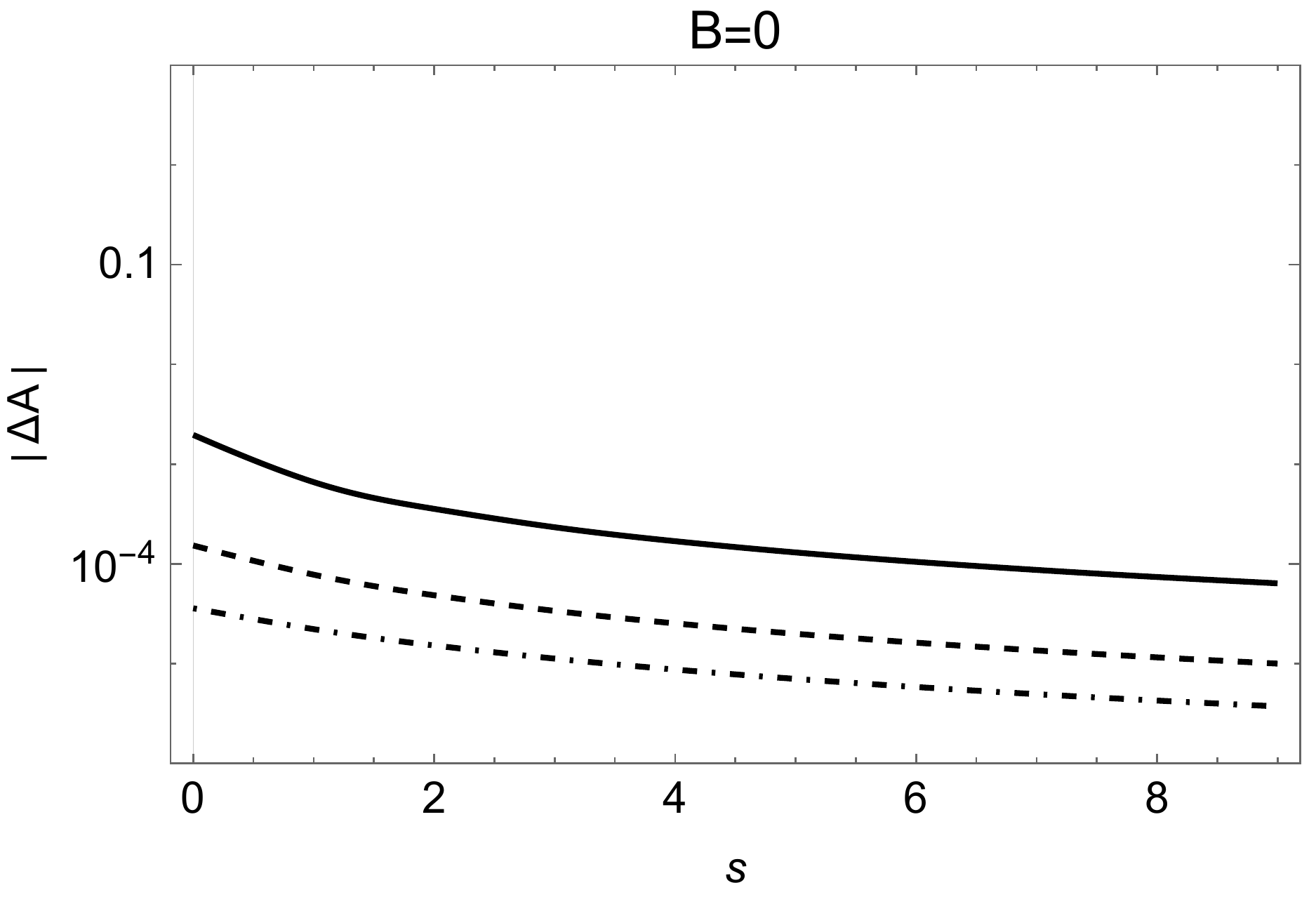}
\endminipage \hspace{0.4cm}
\minipage{0.5\textwidth}
\includegraphics[width=1\textwidth]{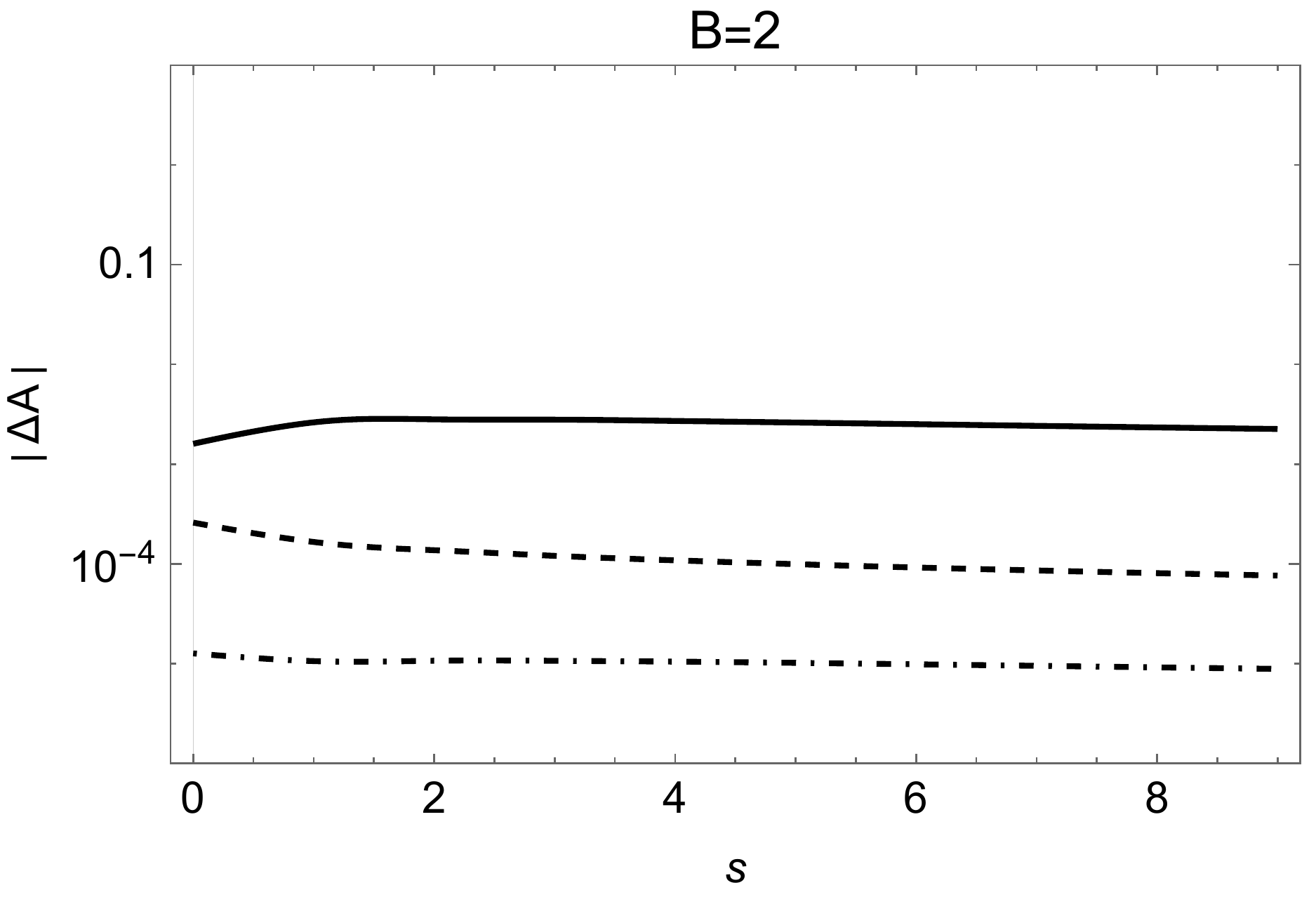}
\endminipage\\
\caption{\small \em The counterpart of Fig.\ref{fig:1} with $j=1$.}
\label{fig:2}
\end{figure}

In order to study the effectiveness of the phase-integral method we 
should now consider more accurately the effects on $A$ of the 
perturbative aspect of the method by taking into account the phase-integral 
quantization condition in Eq. (\ref{(3.14)}). As we have already discussed, 
in the quantization condition we can fix the ``perturbative" order by 
choosing the value of $j$ and varying $s$ in 
the set of natural numbers $\{0,1,2,...\}$. This analysis was
already performed in Ref. \cite{TL1}, where the modulus of 
the difference between the values of $A$, 
$ \left | \Delta A\right | =   \left | A_N-A_{PhI}\right |$, 
obtained at some perturbative order and the 
result of a numerical approach was plotted as a function of $s$. 
We have done the same by considering the fact that our 
numerical value of $A$ was
obtained from the Numerov method - discussed in Sec. \ref{sect:2}. 
The numerical results of the analysis in the previous section 
can be read in Fig. \ref{fig:1} 
where we have collected the plots of $\left | \Delta A\right |$ 
as a function of $s$. The plots are for fixed values of 
$B$ and, for each value of $B$, the solid, dashed and dashed-dotted 
curves correspond to $\ell=0,1,2$, respectively. 
As we can see, for values of $s$ larger than $s=6$ all the curves are smooth; 
the agreement between phase-integral approach and numerical results
is very good for small values of $B$ and it is the best for $\ell=2$, 
and this is true also for $B=5$ and $B=10$ where the case $\ell =0$ and $\ell=1$ shows 
large $\left | \Delta A\right |$. 
In Fig. \ref{fig:2} the plots of $\left | \Delta A\right |$ as a function of $s$ are gathered 
togheter for the same values of $B$ and $\ell$ in Fig. \ref{fig:1}, in this case we are adding,
in our approximation, the contribution of $L^{(3)}$. By comparing the figures it is clear that
for $j=1$ $\left | \Delta A\right |$ is almost two orders of magnitude smaller than the case for 
$j=0$ testifying the quality of the approximation. 

As discussed at the end of the previous section
we have numerically checked our hypothesis regarding the existence of a value of $u_0$
such that, for any solution of the quantization relation, $C(u_{0},m,\alpha^{2})=0$ and 
 ${\rm sn}(u_{0}) \not = 0$, ${\rm cn}(u_{0}) \not = 0$, ${\rm dn}(u_{0}) \not = 0$.

\section{Concluding remarks}
\label{sect:6}
\setcounter{equation}{0}

As far as we can see, the original contributions of our analysis
are as follows.
\vskip 0.3cm
\noindent
(i) An improved evaluation of the phase-integral quantization condition
(3.14) has been obtained, by proving in Eqs. (4.16)-(4.23) that the
$L^{(3)}$ integral contains, for the Coulomb plus linear potential,
the additional term (4.20) whose occurrence was not discussed in Ref. \cite{TL1}.
However, our original analysis has also the merit of showing that
one can explicitly obtain a countable family of complex values of 
$u_{0}$ for which the undesirable additional term in the quantization condition
vanishes.

More precisely, the occurrence of $C$ for arbitrary $u_{0}$ is unavoidable
because the indefinite integrals (4.12)-(4.14) contain the contribution of
Jacobi elliptic functions, and hence the differences in Eq. (4.16) contain 
the effect of $u_{0}$ by virtue of Eqs. (B.1)-(B.10). 
Our detailed presentation makes it possible to verify all 
intermediate steps. Once the order at which
the quantization condition (3.14) is studied is fixed, the value of $u_{0}$
can be obtained , as we have shown in one case at the end of Sec. 4. On going
to the next-to-leading order, one obtains a different value of the appropriate
$u_{0}$, because one is correcting $L^{(1)}$ by adding
$L^{(3)}+L^{(5)}$, or $L^{(3)}+L^{(5)}+L^{(7)}$, ..., and so on. 
In other words, the function $C(u_{0},m,\alpha^{2})$ that pertains to
$L^{(3)}$ differs from the function of the same arguments that pertains
to $L^{(3)}+L^{(5)}$, and so on. Upon setting it to zero one obtains an equation of
increasing difficulty.

\vskip 0.3cm
\noindent
(ii) The Numerov method has been applied successfully to the 
non-relativistic Coulomb plus linear potential. In particular,
the Numerov method has been found to be always in good agreement 
with the early results in Ref. \cite{Eichten}.

We hope that our research will stimulate further work on the 
ultimate nature of asymptotic expansions \cite{P1886,D1980}. 
It also appears of interest to investigate error terms in the
quantization condition (3.14). As is shown in Refs. 
\cite{FR1,Froman1966}, they are of order ${\rm O}(\mu)$, where $\mu$
is defined by the integral
$$
\mu(z,z_{0})=\int_{z_{0}}^{z} \Bigr | q(\zeta) \chi(q(\zeta)) \Bigr | 
d\zeta.
$$
The numerical evaluation of such an integral is a hard task.
Moreover, the experience gained in assessing the phase-integral method
might help in studying the QCD potential approximated in Ref.
\cite{Sumino}. More precisely, at large values of a parameter $N$, the
static QCD potential $V$ consists of four parts corresponding to
$\left \{ {1 \over r}, r^{0}, r^{1}, r^{2} \right \}$ terms, with
logarithmic corrections in the ${1 \over r}$ and $r^{2}$ terms,
and hence one can write, $\rho$ being the dimensionless form 
of $r$ \cite{Sumino}
\begin{equation}
V(\rho,N)=V_{c}(\rho)+ \beta(N)+C \rho + D(\rho,N) +W,
\label{(6.1)}
\end{equation}
where $W$ denotes terms that vanish as $N \rightarrow \infty$, while
in particular
\begin{equation}
V_{c}(\rho) \sim {\pi \over 2 \rho \log(\rho)}
\; {\rm as} \; \rho \rightarrow 0,
\label{(6.2)}
\end{equation}
\begin{equation}
V_{c}(\rho) \sim -{\pi \over \rho} 
\; {\rm as} \; \rho \rightarrow \infty,
\label{(6.3)}
\end{equation}
\begin{equation}
D(\rho,N)=\rho^{2}\left[{1 \over 12}\log(N)+d(\rho)\right],
\label{(6.4)}
\end{equation}
\begin{equation}
d(\rho) \sim -{1 \over 12} \left[2 \log \; \log \left({1 \over \rho}\right)
+\log \left({9 \over 2}\right) +\gamma \right] \; {\rm as} \;
\rho \rightarrow 0,
\label{(6.5)}
\end{equation}
\begin{equation}
d(\rho) \sim -{1 \over 12}\left[2 \log \; \log(\rho)
+\log \left({9 \over 2}\right)+\gamma \right] \; {\rm as} \;
\rho \rightarrow \infty.
\label{(6.6)}
\end{equation}
These logarithmic terms in the potential can be treated by virtue of
the limiting properties obtained from application of
de l'Hopital's rule 
\begin{equation}
\lim_{\rho \to 0} \rho \log(\rho)
=-\lim_{\rho \to 0}\rho=0,
\label{(6.7)}
\end{equation}
\begin{equation}
\lim_{\rho \to 0}\rho^{2}\log \; \log \left({1 \over \rho}\right)
=-{1 \over 2}\lim_{\rho \to 0}{\rho^{2}\over \log (\rho)}=0,
\label{(6.8)}
\end{equation}
but the evaluation of the function $Q$ of Sec. 3 becomes harder,
as far as we can see.

\begin{appendix}
\setcounter{equation}{0}
\section{Jacobi elliptic integrals and elliptic functions}

The complete elliptic integral of first kind,
$K(m)$, can be defined by
\begin{equation}
K(m=k^{2})=\int_{0}^{{\pi \over 2}}{{\rm d} \theta \over \sqrt{1-m \sin^{2}\theta}}
=\int_{0}^{1}{{\rm d}t \over \sqrt{(1-t^{2})(1-k^{2}t^{2})}},
\label{(A1)}
\end{equation}
where $m \in [0,1]$ so that $K(m)$ is real-valued \cite{Abra}.
For the complete elliptic integral of second kind, $E(m)$, one has instead the definition
\begin{equation}
E(m=k^{2})=\int_{0}^{{\pi \over 2}}
\sqrt{1-m \sin^{2}\theta} \; {\rm d}\theta
=\int_{0}^{1}
{\sqrt{1-k^{2}t^{2}}\over \sqrt{1-t^{2}}} \; {\rm d}t.
\label{(A2)}
\end{equation}
Last, the complete elliptic integral of third kind, $\Pi(n,m)$, can
be defined according to
\begin{equation}
\Pi(n,m=k^{2})=\int_{0}^{{\pi \over 2}}
{{\rm d}\theta \over (1-n \sin^{2}\theta)\sqrt{1-k^{2}\sin^{2}\theta}}.
\label{(A3)}
\end{equation}
Note that, on defining
\begin{equation}
u=\int_{0}^{\phi}{{\rm d}\theta \over \sqrt{1-m \sin^{2}\theta}},
\label{(A4)}
\end{equation}
one finds
\begin{equation}
u \Bigr(\phi={\pi \over 2},m=k^{2} \Bigr)=K(m=k^{2}).
\label{(A5)}
\end{equation}
The Jacobi elliptic functions that we use in Sec. $3$ are
defined from Eq. (A4) according to
\begin{equation}
{\rm sn}(u)=\sin \phi, \;
{\rm cn}(u)=\cos \phi, \;
{\rm dn}(u)=\sqrt{1-m \sin^{2}\phi},
\label{(A6)}
\end{equation}
where the angle $\phi$ is said to be the amplitude:
$$
\phi={\rm arcsin}({\rm sn}(u)).
$$
The work in Ref. \cite{TL1} denotes by $K(k)$ what we denote
by $K(m=k^{2})$, and the same remark holds for the other two
elliptic integrals. 

\section{Evaluation of $L^{(3)}$}
\setcounter{equation}{0}

In the course of obtaining Eq. (4.16), we have used Eqs. (4.12)-(4.14)
and the identities
\begin{equation}
{\rm cn}(u+v)={{{\rm cn}(u) \; {\rm cn}(v)-{\rm sn}(u) \; {\rm sn}(v) 
\; {\rm dn}(u) \; {\rm dn}(v)} \over \Bigr(1-m \; {\rm sn}^{2}(u) \;
{\rm sn}^{2}(v)\Bigr)},
\label{(B1)}
\end{equation}
\begin{equation}
{\rm dn}(u+v)={{{\rm dn}(u) \; {\rm dn}(v)-m \; {\rm sn}(u)
\; {\rm sn}(v) \; {\rm cn}(u) \; {\rm cn}(v)}\over
\Bigr(1-m \; {\rm sn}^{2}(u) \; {\rm sn}^{2}(v)\Bigr)},
\label{(B2)}
\end{equation}
\begin{equation}
{\rm sn}(u+v)={{{\rm sn}(u) \; {\rm cn}(v) \; {\rm dn}(v)
+{\rm sn}(v) \; {\rm cn}(u) \; {\rm dn}(u)} \over
\Bigr(1-m \; {\rm sn}^{2}(u) \; {\rm sn}^{2}(v)\Bigr)},
\label{(B3)}
\end{equation}
\begin{equation}
{\rm cn}(K)=0, \; {\rm dn}(K)=\sqrt{1-m}, \; {\rm sn}(K)=1,
\label{(B4)}
\end{equation}
which imply that
\begin{equation}
{\rm cn}(u_{0}+K)=-\sqrt{1-m} \; {{\rm sn}(u_{0})\over {\rm dn}(u_{0})},
\label{(B5)}
\end{equation}
\begin{equation}
{\rm dn}(u_{0}+K)={\sqrt{1-m}\over {\rm dn}(u_{0})},
\label{(B6)}
\end{equation}
\begin{equation}
{\rm sn}(u_{0}+K)={{\rm cn}(u_{0})\over {\rm dn}(u_{0})},
\label{(B7)}
\end{equation}
and hence
\begin{equation}
{{\rm cn}(u_{0}+K) \; {\rm dn}(u_{0}+K) \over {\rm sn}(u_{0}+K)}
=(m-1){{\rm sn}(u_{0})\over {\rm cn}(u_{0}) \; {\rm dn}(u_{0})},
\label{(B8)}
\end{equation}
\begin{equation}
{{\rm dn}(u_{0}+K) \; {\rm sn}(u_{0}+K)\over {\rm cn}(u_{0}+K)}
=-{{\rm cn}(u_{0})\over {\rm dn}(u_{0}) \; {\rm sn}(u_{0})},
\label{(B9)}
\end{equation}
\begin{equation}
{{\rm cn}(u_{0}+K) \; {\rm sn}(u_{0}+K) \over 
{\rm dn}(u_{0}+K)}=-{{\rm cn}(u_{0}) \; {\rm sn}(u_{0}) \over
{\rm dn}(u_{0})}.
\label{(B10)}
\end{equation}
 
\end{appendix}

\section*{Data availability statement}
The datasets generated during the current study are available from the
corresponding author on reasonable request.

\end{document}